\documentstyle[eqsecnum,aps,twocolumn,psfig,myabstract]{revtex}
\begin{document}
\draft

\def\th{\thinspace}
\def\qquad{\quad\quad}
\def\ngth{\negthinspace}
\def\Teff{{$T_{e\!f\!f} $}}

\title{{\it submitted to} {\rm ASTROPHYSICAL JOURNAL LETTERS, Aug. 1997}
\break
\ \ \break
On the Deficiency of 8--10 Day Galactic Cepheids}
\author{J. Robert Buchler}
\address{Physics Department, University of Florida, Gainesville, FL 32611}
\author{Marie-Jo Goupil}
\address {Observatoire de Paris, Meudon}
\author{Rick Piciullo}
\address{Physics Department, University of Florida, Gainesville, FL 32611}


\address{\ }
\myabstract{
 The Galactic Cepheid period histogram has a strong dip between 8 and 10 days
that has defied an explanation based on evolutionary and linear pulsation
studies.  We show here that this deficiency is caused by the instability of the
{\sl nonlinear} fundamental pulsation cycle in this period range.  The strong
metallicity dependence of this instability is consistent with the absence of a
corresponding minimum in the Magellanic Cloud data.  Our results also suggest
that the Galactic Cepheids must have a large spread in metallicity.
 }
\address{\ }




\maketitle

\narrowtext

It is a well known fact that the observed period distribution of the Galactic
Cepheids has a pronounced minimum in the 8--10\th d period range, as shown in
Fig.~1 which displays the histogram for the period distribution of the Galactic
Cepheids constructed from the Galactic Cepheid Database of Fernie et
al. (1995).  Since it is difficult to separate overtone and fundamental
pulsators for the Galaxy above 5\th d our histogram necessarily contains both
the fundamental and the overtone Cepheids.

Becker, Iben \& Tuggle (1977) combined the location and duration of Cepheid
model crossings of the instability strip from evolutionary calculations with a
birth-rate function to infer a theoretical period histogram.  One of their
conclusions was that a minimum in the distribution was not compatible with a
standard birth-rate function, and could only be explained if an ad hoc
two-component birth-rate function were adopted (cf. however Chiosi 1989).
Their evolutionary computations were performed with the now superseded Los
Alamos opacities which are now known to be considerably too weak.  However,
while the new opacities will produce a different period distribution it is
difficult to see how they might cause a two-humped one.

In this Letter we show that the minimum in the period distribution of Cepheids
is a result of the nonlinear dynamics associated with the fundamental
pulsations of the Cepheids, and that it has therefore nothing to do with
evolutionary calculations.

The proper mass--luminosity (ML) relations for the Cepheid model sequences
would normally have to be obtained from evolution calculations.  However, at
the present time there is enough uncertainty and disagreement (Buchler et
al. 1996, Beaulieu et al. 1997) to lead us to determine these ML relations
differently.  From the structure of the Fourier decomposition parameters
$\phi_{21}$ and $R_{21}$ for the Galaxy and for the Magellanic Clouds one
infers that the 2:1 resonance occurs in the vicinity of a period of 10\th d.
Based on this fact we construct the ML relation as follows: For each
metallicity $Z$ we determine the mass $M$ and luminosity $L$ for which the
$P_0$=10\th d equilibrium-model is resonant viz. $P_2/P_0$=1/2 (bump Cepheid)
with an effective temperature \Teff\ that lies $\Delta T$=100\th K degrees to
the right of the fundamental blue edge. \th (Mathematically, for given
composition parameters $X$ and $Z$ and a given $\Delta T$ we solve for
$P_0(M,L,T_{ef\!f})$=10, $P_2(M,L,T_{ef\!f})$=5 with Teff\ = $T_{BE}$--$\Delta
T$ where $T_{BE}=T_{BE}(M,L)$).  From these anchor values ($M$, $L$) we then
derive a ML relation with a slope chosen to be 3.56 that is close to what
evolutionary calculations indicate.

\psfig{figure=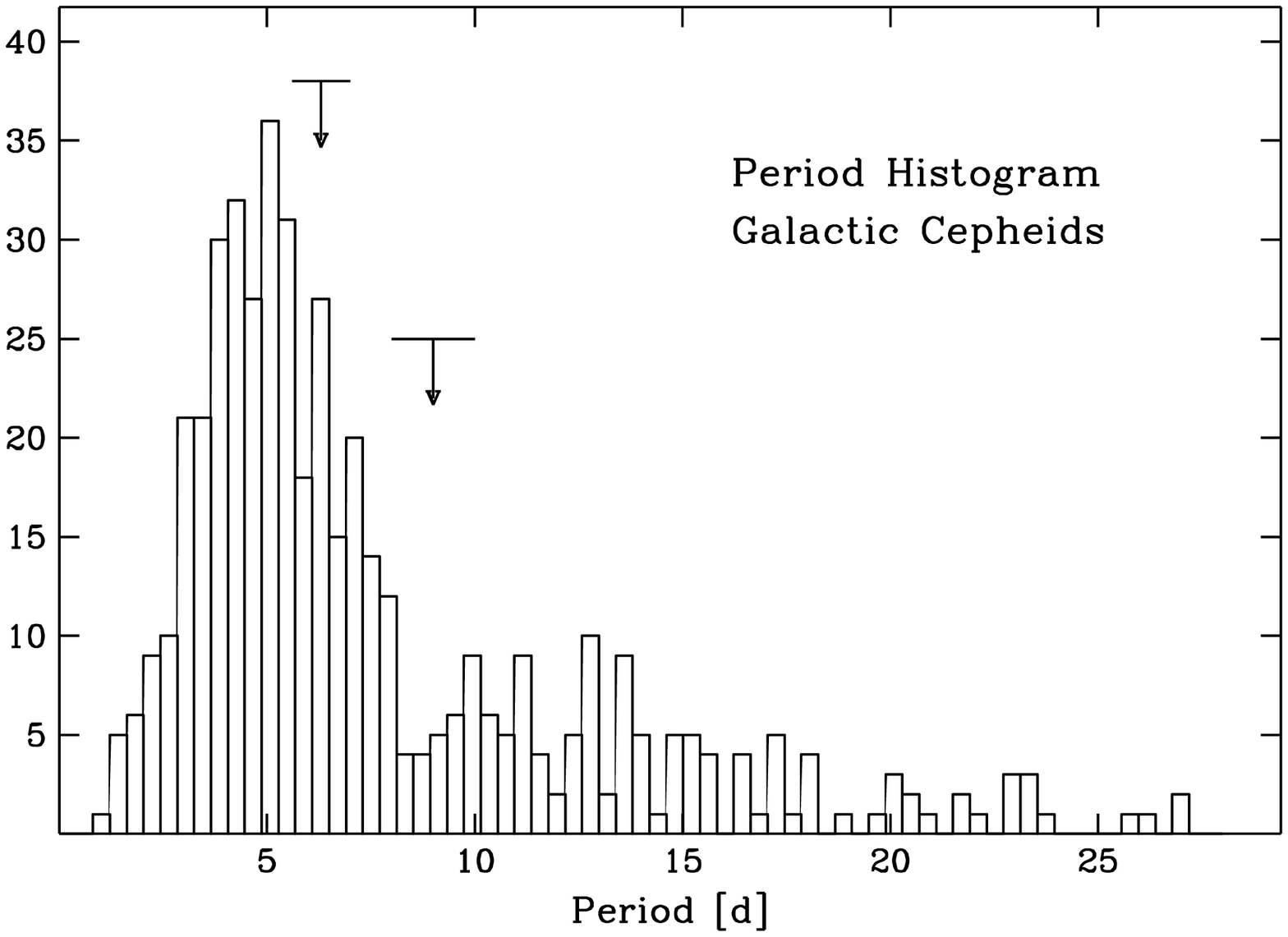,width=3.5in}

\noindent {\bf Fig.1:} Period distribution of fundamental and overtone Galactic
Cepheids d'apr\`es Fernie et al.
\vspace{10pt}

We use the Livermore OPAL95 opacities (Iglesias \& Rogers 1996) combined with
the molecular opacities of Alexander \& Ferguson (1994), to compute the
nonlinear fundamental pulsations with the relaxation method of Stellingwerf
(1974).  We put a temperature anchor at 11,000\th K with 30 constant mass
shells in the surface, and with 90 geometrically increasing zones inward.  The
pseudo-viscosity parameters are $C_Q$=4 and $\alpha$=0.01.  Convection is
ignored, and we caution that the models lose their validity far from the blue
edge.  Concomitantly with the relaxation to the periodic pulsation the code
performs a Floquet analysis of the limit cycles (i.e. the periodic finite
amplitude pulsations) (e.g. Buchler 1990).  We recall that the Floquet
exponents measure the linear stability of the limit cycle to perturbations with
respect to all the possible modes (e.g. Ince 1944): For a limit cycle to be
stable, and thus to be observable, {\sl all} Floquet exponents, $\lambda_k$,
have to be negative.  This stability analysis is an extremely useful byproduct
of the relaxation method.  In fact without the Floquet analysis we might have
to integrate thousands of cycles to ascertain stability of a limit cycle, and,
in the case of instability, we would not be able to compute the limit cycles at
all and determine {\sl how} unstable they are.  The relaxation code on the
other hand is robust enough to converge on a limit cycle even when the latter
is mildly unstable.

\psfig{figure=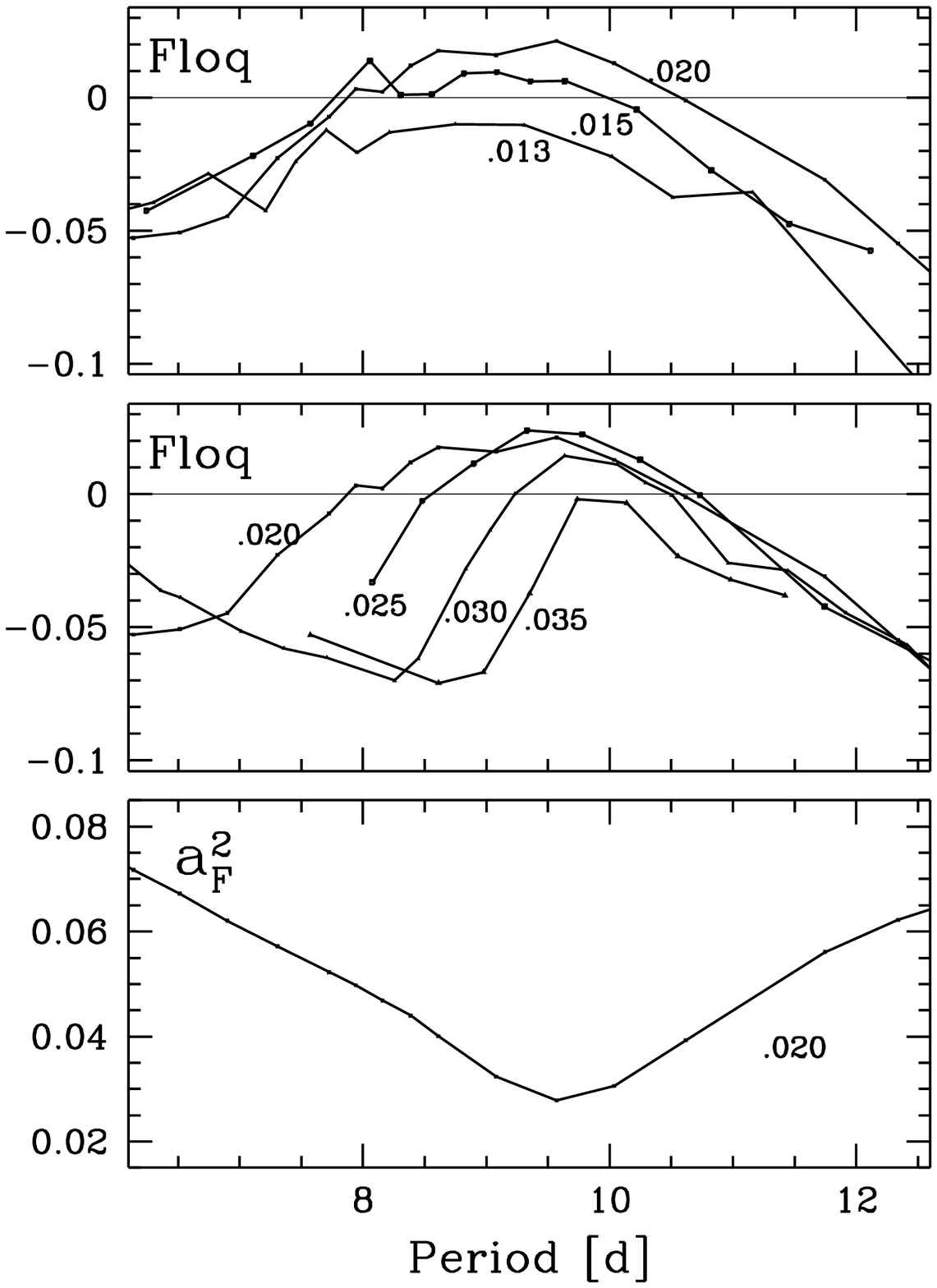,width=3.2in}
 \noindent {\bf Fig.2}: {\sl Top, middle:} Floquet stability exponent
$\lambda_1$ for the nonlinear fundamental Cepheid pulsations as a function of
period and metallicity $Z$ (with $X$=0.70) for models 100\th K to the right of
the fundamental blue edge; {\sl Bottom:} Square of the {\sl radius} Fourier
amplitude of the limit cycle scaled by the period, $a_F$, cf text.
\vspace{10pt}

In Fig.~2 we exhibit the behavior of the Floquet stability exponent,
$\lambda_1$, of the fundamental limit cycle as a function of its (nonlinear)
period $P^{nl}_0$.  (In this period region it is the perturbation with the
first overtone that is the least stable).  The six sequences have
metallicities, with $Z$ ranging from 0.013 to 0.035.  In order to avoid
cluttering we have split the figure into two subfigures, the top showing the
metallicities $Z$= 0.013, 0.015 and 0.020, and the bottom the values 0.020,
0.025, 0.030 and 0.035.  The hydrogen mass fraction has been chosen at
$X$=0.70.  All model sequences run $\Delta T$=100~K to the right of the
fundamental blue edge.  For metallicities in the range $Z$ $\approx$ 0.014 to
0.035 these fundamental limit cycles are thus unstable to a perturbation in the
first overtone.

In order to see the effect of the location of the Cepheids with respect to the
blue edge we display in Fig.~3 results obtained for a sequence with $\Delta T$
= 400~K \th (with the same ML relation as for the 100~K sequences described
above).  The width of the period region of unstable limit cycles thus depends
somewhat on location with respect to the instability strip, and shrinks with
$\Delta T$.

\psfig{figure=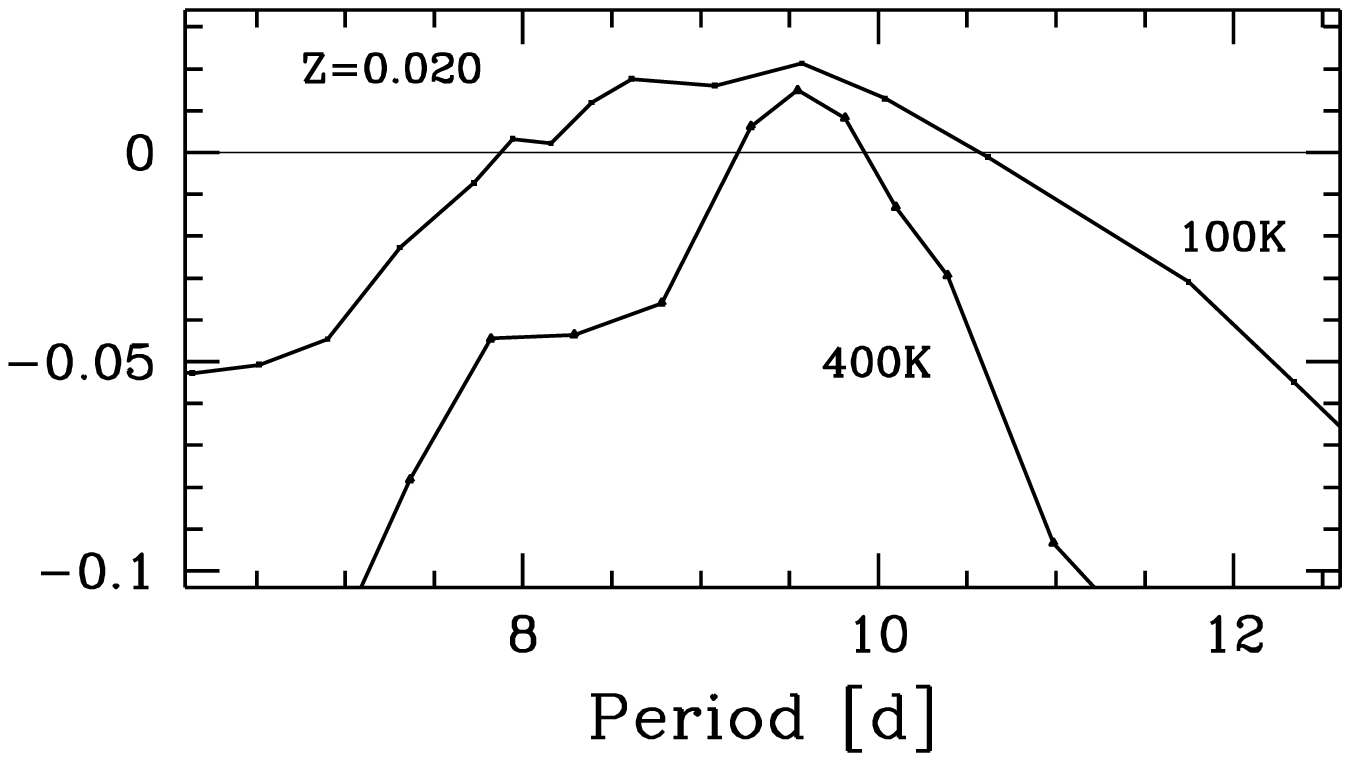,width=3.1in}
 \noindent {\bf Fig.3}: Floquet stability exponent $\lambda_1$ for the
nonlinear
fundamental Cepheid pulsations as a function of period for two sequences with
$\Delta T$ = 100 and 400\th K and with $Z$=0.020 and $X$=0.70.
\vspace{10pt}

Older nonlinear Cepheid pulsations that were performed with the now obsolete
Los Alamos opacities (Moskalik \& Buchler 1991) indicated that the fundamental
Cepheid pulsations were stable in the vicinity of 10 d (except for an
absolutely minute range, as their Fig.~1 shows).  It is quite clearly the
overall increase in the opacities that is responsible for the instability.  One
can get a good feeling for the sensitivity of the Floquet exponents to opacity
from Fig.~3 of Buchler (1996) where the results of calculations with the
OPAL95, the OPAL93 and the Los Alamos opacities are compared.  Roughly
speaking, the change from OPAL93 to OPAL95 is equivalent to an increase of $Z$
from 0.02 to 0.03, for example (Buchler, Koll\'ath, Beaulieu \& Goupil 1996).

We caution the reader though that there is some sensitivity of the Floquet
exponents to the zoning and to other numerical parameters that are used in the
nonlinear calculations (Kov\'acs 1990, Yecko, Koll\'ath \& Buchler 1997), as
well as to convection whose effects we have ignored.  The precise values of the
metallicity $Z$ for the onset of instability and of the period range thereof
should therefore not be taken too literally, although the {\sl existence} of an
instability in the broad 8--10\th d range seems to be essentially independent
of such numerical parameters.

\vskip 10pt

 What is the reason behind the instability of the fundamental limit cycles?  We
show now that the resonance between the fundamental mode of oscillation and the
second overtone ($P_2/P_0=1/2$) cause an overall decrease of the pulsation
amplitude that in turn destabilizes the fundamental limit cycle.  We note that
this is the same resonance that causes the familiar Hertzsprung bump
progression.

It is a well known observational fact that the pulsation amplitudes of the
classical Cepheids exhibit a substantial drop for periods in the vicinity of
10\th d.  The same feature is found in numerical modelling.  The amplitude
equation formalism (e.g. Buchler 1993) explains how nonlinear effects
associated with the resonance $P_2/P_0$=1/2 between the fundamental and the
second overtone are responsible for the drop in the overall pulsation
amplitude.  Furthermore, large amplitudes increase the stability of a limit
cycle pulsation, as the same formalism shows.  Indeed the Floquet exponent
$\lambda_1$ behaves as (Buchler, Moskalik \& Kov\'acs 1991)
 $$\lambda_1 = (\kappa_1 - q_{10} A_0^2 - q_{12} A_2^2 ) \th P_0 
 \eqno(1)$$
 where $\kappa_1$ is the linear growth-rate of mode $1$, $q_{1j}$ are nonlinear
coupling coefficients that depend on the structure of the star, and $A_j$ are
the pulsation amplitudes of the excited modes, i.e. the fundamental and the
resonant second overtone here.

For simplicity, let us make a few approximations.  First we ignore the smaller
amplitude of the second overtone, $A_2$ compared to $A_0$.  Second, with good
reasons we assume that the cubic coupling coefficients $q$ are positive and
that they scale as $q_{10}=c_{10} P_0$ (Kov\'acs \& Buchler 1989).  Third we
disregard the small variation of $\kappa_1 P_0$ over the plotted range of
periods.  In Eq.~1 the amplitudes are relative, i.e. they refer to $\delta
R(t)/R$.  Thus we associate $A_0$ with the lowest Fourier amplitude $a_1$ of
the absolute computed radius variations, scaled by the stellar radius $R_*$,
i.e. with $a_F=a_1/R_*$.  How the radius scales with period can be found by the
following rough estimate.  Using the mass--luminosity relation $L \sim
M^{3.56}$, the period--mean density relation $P_0\sim \rho^{-1/2}\sim
R^{3/2}M^{-1/2}$, the surface luminosity equation $L\sim R^2 T^4$ and the shape
of the blue-edge $T \sim L^\alpha$, with $\alpha\sim -0.05$, one finds
 $$P_0 \sim R^\beta\quad , \quad\quad
 \beta={3\over 2} - {1\over 3.56 (1-4\alpha)} \sim 1.27$$
 With $\beta\approx 1$ Eq.~1 is reduced to
 $$\lambda_1
 \approx \kappa_1 \th P_0 - c_{10} \th a_F^2
 \eqno(2)$$
 A comparison of the bottom panel of Fig.~2 with the upper panels
clearly confirms the correlation between pulsation amplitude and stability.

\vskip 10pt

On the basis of the stability properties of the fundamental limit cycles
exhibited in Figs.~2 and 3 we thus reach the following conclusions.  First, for
metallicities in the range $Z$ $\approx$ 0.014 to 0.035 the fundamental limit
cycle is unstable to a perturbation in the first overtone.  Second, the window
of instability shifts to higher period with increasing $Z$.

Turning now to the astronomical implications we note that the Galactic period
histogram has a dip, but not an actual gap in the 8--10\th d range as it should
have if the metallicity of all Cepheids were as large $Z$=0.02.  If the minimum
of the histogram is now interpreted as a dispersion in metallicity the observed
$\approx$ 8--10\th d fundamental Cepheids, approximately a third, must
therefore have lower metallicity, $Z<$ 0.013 \th (or else an unlikely
$Z>$0.035).  We note though that Becker et al. (1977), albeit on totally
different grounds, also suggested that Galactic metallicity dispersion is large
(a range $Z_{max}/Z_{min}$ of 3 to 5).

However, it may be objected that since the stability of the fundamental limit
cycles depends on the location with respect to the instability strip a wide
strip could also reduce the Cepheid deficiency especially on the lower period
side.  There are evolutionary arguments (Becker et al. 1977) that the Cepheid
instability strip may be much narrower than generally assumed.  An estimation
of the instability strip from the Fourier decomposition parameters also
suggested a narrower rather than larger instability strip (Buchler, Moskalik \&
Kov\'acs 1990).

How much of the survival of 8--10\th d Cepheids is indeed due to a dispersion
in metallicity rather than due to the width of the instability strip can be
tested with an observational measurement of the metallicities of the individual
Cepheids.

\vskip 5pt

While there is a deficiency of fundamental Cepheid pulsators in the 8--10\th d
range, there is absolutely no reason to believe that there is a deficiency of
the corresponding stars.  As expected, hydrodynamical calculations show that
these stars pulsate in stable first overtone limit cycles.  They should
therefore show up as an excess in the histogram at lower periods (lower by a
factor $\approx$ 0.7, the typical period ratio $P_1/P_0$).  Indeed, the
histogram of Fig.~1 is certainly compatible with an excess in the 5.6--7.0\th d
range.  (The left overbar shows the first overtone period range corresponding
to the fundamental range shown on the right.)

\vskip 10pt

Turning now to other galaxies, we recall that Becker et al. (1977) present also
a period histogram for M31, a galaxy that also has a relatively high
metallicity.  Again the histogram shows the dip in the 8--10\th d range,
consistently with our results.

In contrast, the Magellanic Clouds are known to have considerably lower average
metallicities than the Galaxy, and indeed, in agreement with our stability
analysis, the period histograms for the Magellanic Clouds (Becker et al 1977)
do not show much indication of a minimum.  The small hollow near 9\th d, if
statistically significant, may again be an indication of a spread in
metallicity putting some Magellanic Cepheids above the threshold value of
$Z\approx$ 0.013.

\vskip 5pt

If the 2:1 resonance at 10\th d plays such an important role one may wonder if
other resonances might make themselves felt, especially because the Galactic
Cepheid period distribution (Fig.~1) seems to indicate a deficiency of Cepheids
in two other places.

First, if the gap near $P\approx$ 25\th d is indeed significant it has an
interesting implication.  Numerical hydrodynamic calculations (Moskalik \&
Buchler 1991 and Moskalik, Buchler \& Marom 1992 for the new opacities) show
that the $P_0/P_1 = 3/2$ resonance can destabilize the fundamental Cepheid
pulsation and lead to a {\sl periodic} pulsation with double period, in which
alternating cycles differ only slightly however.  We note though that Fernie
(preprint) has not found any evidence of alternations in his Cepheid data.
However, in support of this scenario Antonello \& Morelli (1996) has suggested
that the star CC~Lyr does exhibit alternations.  The observational data of
CC~Lyr are very limited though, and additional observations of this star are
necessary to confirm the existence of alternations.  Is it possible that the
dip in Fig.~1 could arise because such stars with alternating cycles may not
have been classified as Cepheids?  If this resonance is indeed strong enough to
cause instability to alternations then this would add an additional
observational constraint for Cepheid models.

 Second, Fig.~1 perhaps also suggests a minimum in the vicinity of 19 d.  One
notes that this is the period region where the fundamental mode is in a 4:1
resonance with the fifth overtone ($P_5/P_0$=1/4).  However, the linear
stability of the 4th overtone is quite large, and a hydrodynamical survey is
necessary to verify whether it can play a sufficiently large dynamical role to
destabilize the fundamental limit cycle.

\vskip 10pt

In conclusion, the simple and natural explanation for the deficiency of
$\approx$\th 8--10\th d Cepheid variables is that the fundamental mode limit
cycles are unstable.  The corresponding stars pulsate in the first overtone
with period $P_1\approx 0.7 P_0$, giving rise to a relative excess of overtone
Cepheids in the corresponding range of $\approx$\th5.6--7.0\th d periods.  It
is thus not necessary to invoke an ad hoc two-component birth-rate function to
explain the Cepheid period distribution.

Our calculations and the agreement they provide with the minimum in the
observed Galactic Cepheid distribution, and the lack of a minimum in the lower
metallicity Large Magellanic Cloud data also provide a further confirmation
that the new opacities are in the right ball-park.  Indeed, if they were much
weaker, the 8--10\th d fundamental Cepheid pulsations would all be stable, thus
not giving the observed minimum.  On the other hand if they were much stronger,
then they would predict a minimum for the LMC as well, and a smaller one, or
none at all for the Galactic Cepheids, in contrast with observation.

\vspace{10pt}

It is a pleasure to thank Zoltan Koll\'ath, Jean-Philippe Beaulieu and Phil
Yecko for fruitful conversations.
 This research has been supported by the NSF (AST95--18068, INT94--15868) at
UF, by the CNRS (D03350/17) at DASGAL and an RCI account at the NER Data Center
at UF. 
 This paper was completed in the stimulating atmosphere of the Aspen Center for
Physics.

\end{document}